\documentclass[11pt]{article}

\usepackage{epsf}
\usepackage{graphicx}
\usepackage{amssymb}
\usepackage{amsmath}

\newcommand{\vs}[1]{\rule[- #1 mm]{0mm}{#1 mm}}

\begin{document}

\renewcommand{\thefootnote}{\fnsymbol{footnote}}

\vs{10}
\rightline{LAPTH-Conf-1164/06\qquad}

\begin{center}

{\Large {\bf{PHOTON AND DILEPTON PRODUCTION}}}\\
\vspace{.2cm}
{\Large {\bf{IN HEAVY ION COLLISIONS AT LHC}}}
\footnote{Talk given at  
the 2nd International Conference on Hard and Electromagnetic Probes of
High-Energy Nuclear Collisions, June 9-16 2006, Asilomar, Pacific Grove,
California}\\
\vspace{1.cm}

{\large P. Aurenche}

{\em Laboratoire d'Annecy-le-Vieux de Physique Th\'eorique LAPTH,
\footnote{UMR5108 du CNRS associ\'ee \`a l'Universit\'e de Savoie.}\\
B.P. 110, F-74941 Annecy-le-Vieux Cedex, France} \\

\end{center}

\centerline{ \bf{Abstract}}
\noindent
We review various production mechanisms of photons and small mass dileptons
at large transverse momentum in heavy ion collisions at the LHC. Their relevance as a signal for
quark-gluon plasma formation is discussed.

\centerline{\bf
---------------------------------------------------------------------}

\vs{2}

It is commonly accepted that direct photons, once emitted in a heavy ion
collision, do not interact with the hot and dense matter produced. Photons are
radiated during all stages of the collision and therefore they tell us about
the history of the collision and eventually about the hot matter (quark-gluon
plasma) formation. This is unlike hadrons which reflect the physics after the
plasma has cooled down. However the radiative decays of hadrons provide a very
large background to direct photons, specially in the lower $p_T$ range of the
spectrum. To avoid the largest such background ($\pi^0 \rightarrow \gamma
\gamma$) it is interesting to look at the production of small mass lepton
pairs, which involve the same dynamical processes as real photons when the
ratio mass over momentum is small. In the following, I discuss the transverse
momentum spectrum of real photons and small mass virtual photons. Correlation
functions involving a large energy photon are also considered as they are shown
to give a detailed probe of the jet energy loss mechanism.

There are several production mechanisms of real or virtual photons in heavy ion
collisions~\cite{yellowrep}. In {\em prompt} processes photons are radiated in
the interaction of quarks and gluons of the incoming nucleons: their rate
involves the parton densities in the colliding ions. They lead to a power
damped spectrum which should dominate at high momentum. {\em Thermal} photons
are produced in the hot quark or hadronic matter formed during the collision:
their rate is directly related to the temperature of the hot matter. Finally,
{\em mixed} processes involve a prompt high energy quark colliding with a
parton in the medium. We discuss each mechanism, commenting on the
accuracy of the theoretical calculations.

\begin{figure}[htbp]
\begin{center}
\includegraphics[width=15cm,height=9cm]{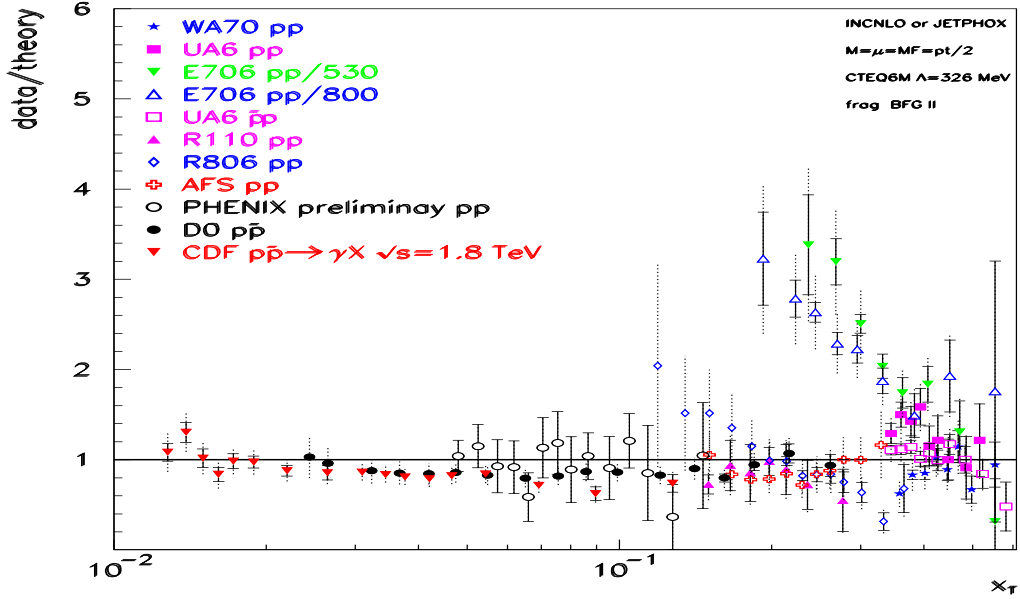}
\end{center}
\caption{Prompt photon production in $pp$ and $p \bar p$ collisions (from
\cite{us2}).}
\label{figure1} 
\end{figure}

\begin{figure}
\begin{center}
\includegraphics[width=13cm]{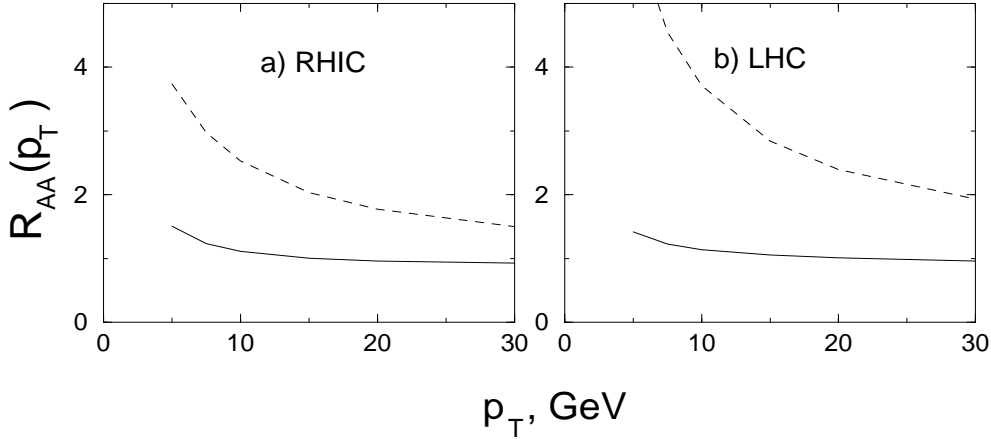}
\end{center}
\caption{Ratio of direct photon production in $AA$ collisions compared to $pp$
collisions. Dashed line: taking into account only the induced emission by
the medium. Solid line: taking into account induced emission as well as jet
energy loss effects. The figure is from Zakharov~\cite{zakharov}.}
\label{figure2}
\end{figure}

Prompt processes are under good control in $pp$ collisions where many data
sets, from about 20 GeV to 2 TeV, exist~(fig.~\ref{figure1}). The underlying
processes, direct and bremsstrahlung, are calculated in the next-to-leading
logarithm (NLO) accuracy~\cite{us1,us2}. The data cover a large range in $x_T$, down
to 10$^{-2}$. All data sets but one (E706) agree, within the error bars, with
the theoretical predictions using standard NLO structure~\cite{cteq} and
fragmentation~\cite{guillet-frag} functions. Extrapolating to LHC at 5.5 TeV
and for $p_T$ values of interest requires controling the theory down to $x_T
\sim 10^{-3}$: in such a range predictions become uncertain because the
bremsstrahlung process becomes dominant, involving the essentially
unconstrained gluon into photon fragmentation function. Another issue is the
validity of the NLO calculation in that domain where resummed calculations of
the single inclusive spectrum are called for.\\
In heavy ion collisions, the basic structure of prompt processes remains the
same except for modifications of the parton distributions due to nuclear
effects, and of
the fragmentation functions due to the interaction of the final hard parton
with the hot medium. Several parametrisations of nuclear structure functions
exist which differ in the amount of shadowing affecting the gluon at small
$x$~\cite{eks,hkm,deflorian}. This will directly modify the photon rate which is
proportional to the gluon structure function. Concerning final state effects,
the mechanism of jet energy loss has been much studied~\cite{energyloss} . It
affects only photons produced by bremsstrahlung. In the first parametrisations
used it was  taken for granted that the same mechanism as for hadron production
was at work, namely it was implicitly assumed that the photon was produced
outside the plasma volume and the production rate was therefore reduced.
More recently, however, it was emphasized~\cite{zakharov} that final state
interaction of the fast quark with the medium  induced photon radiation: in
that case the photon is emitted from within the hot plasma. Combining the jet
energy loss mechanism and the enhanced photon emission will more or less
compensate as shown in the model calculation in fig.~\ref{figure2}. It
should be noted that, in an {\em a priori} NLO calculation of the photon
spectrum, thermal modifications of the fragmentation functions affect the
delicate scale compensation mechanism occuring between the lowest order term
and the higher order corrections, so that QCD calculations of this mechanism
become of leading order accuracy only in heavy ion collisions.

The next production mechanisms we consider are purely thermal. They are usually
calculated in the framework of the effective theory of Braaten and
Pisarski~\cite{braaten} where hard thermal loops are resummed. Two classes of
processes are contributing. The first one involves Compton ($Gq \rightarrow
\gamma q$) and annihilation ($q \bar q \rightarrow \gamma G$)
processes~\cite{oneloop-phot,oneloop-lep}. The second one is of bremsstrahlung type ($G q
\rightarrow \gamma G q$ and $q q \rightarrow \gamma q  q$) together with the
related crossed processes of type  $G q \bar q \rightarrow \gamma G$ or $q q
\bar q \rightarrow \gamma q$~\cite{twoloop}. In these cases all initial partons
are thermal so that the calculated rate of production of a real photon is
exponentially damped in the energy. For a photon of momentum $(E, {\bf p})$ the
calculated rate is of the form $E dN / d {\bf p} \sim  
\alpha~\alpha_s~\exp{(-E/T)}~T^2~f(E/T)$ where $f$ is a soft function of $E/T$.
To obtain the leading order result, the second class of processes requires the
resummation of ladder diagrams in the effective theory~\cite{moore,gelis}, which is
equivalent to taking into account the Landau-Pomeranchuk-Migdal effect or
multiple scattering effects in the plasma. Thermal photons are also produced in
the hot hadronic stage following the QGP phase~\cite{rapp}.
 \begin{figure}
\begin{center}
\includegraphics[width=14cm]{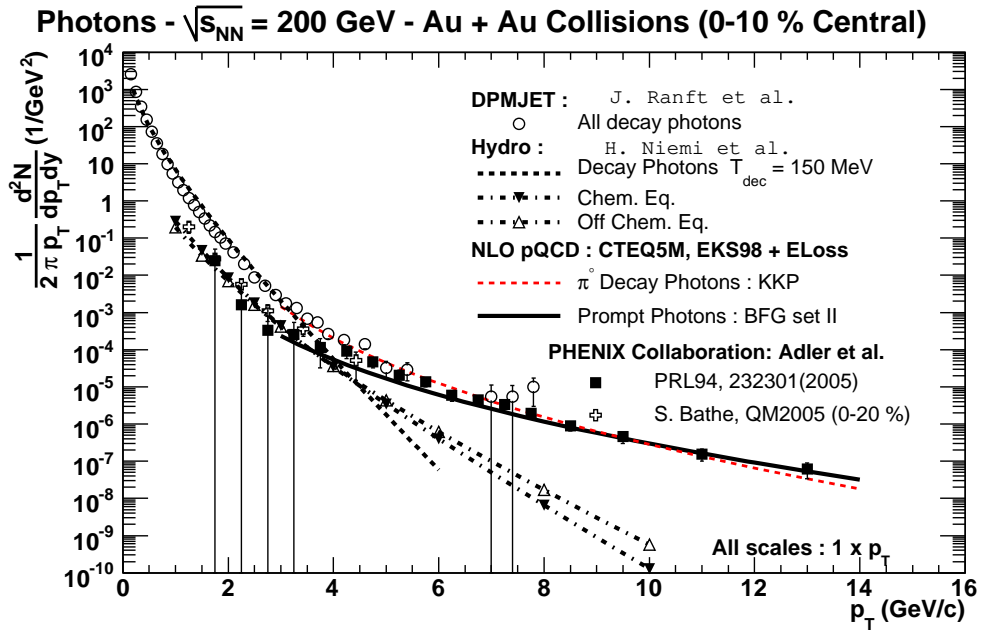}
\end{center}
\caption{Single photon spectrum in $Au Au$ collisions at RHIC
(from~\cite{yellowrep}). Data based on real photon~\cite{phenix} (black
squares) and lepton pair~\cite{phenix-dilep} (open crosses) measurements are
added for comparison with the predictions ({\em courtesy H. Delagrange}).}
\label{figure3}
\end{figure}
In order to make predictions and compare with real data it is necessary to
implement these rates in an hydrodynamical code which describes the expansion
and cooling of the plasma and the transition to the hadronic phase. In the
applications below we use the code of Ruuskanen {\em et
al.}~\cite{ruuskanen}. The important parameters are constrained from
specified initial conditions and the predictions for various hadronic spectra
were shown to agree  with the measurements in $Au Au$ collisions at RHIC. In
fig.~\ref{figure3} the thermal (dash-dotted lines) and prompt
(solid line) processes discussed above are shown and compared to PHENIX
data~\cite{phenix}. As expected, one sees the dominance of thermal processes at
the smaller $p_T$ values. The background ('decay photons') is estimated using
two completely different models, the NLO + hydrodynamic model as well as the
DPMJET model of Ranft {\em et al.}~\cite{ranft} involving the production and
decay of hadronic strings. Both agree. It is worthwhile noting the agreement of
the theoretical predictions for direct photons (prompt + thermal photons
model)~\cite{yellowrep} with the published data~\cite{phenix,phenix-dilep}
within the large error bars.
\begin{figure}
\begin{center}
\includegraphics[width=14cm]{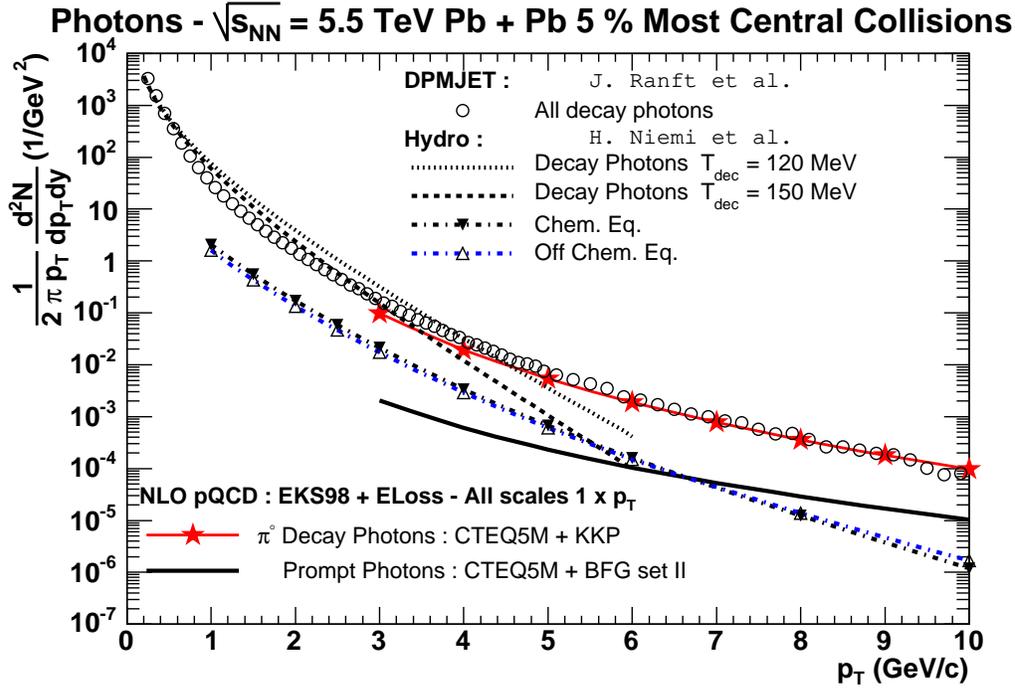}
\end{center}
\caption{The various components of the single photon $p_T$ spectrum in $Pb Pb$
collisions at LHC.}
\label{figure4}
\end{figure}
\begin{figure}
\begin{center}
\includegraphics[width=14cm]{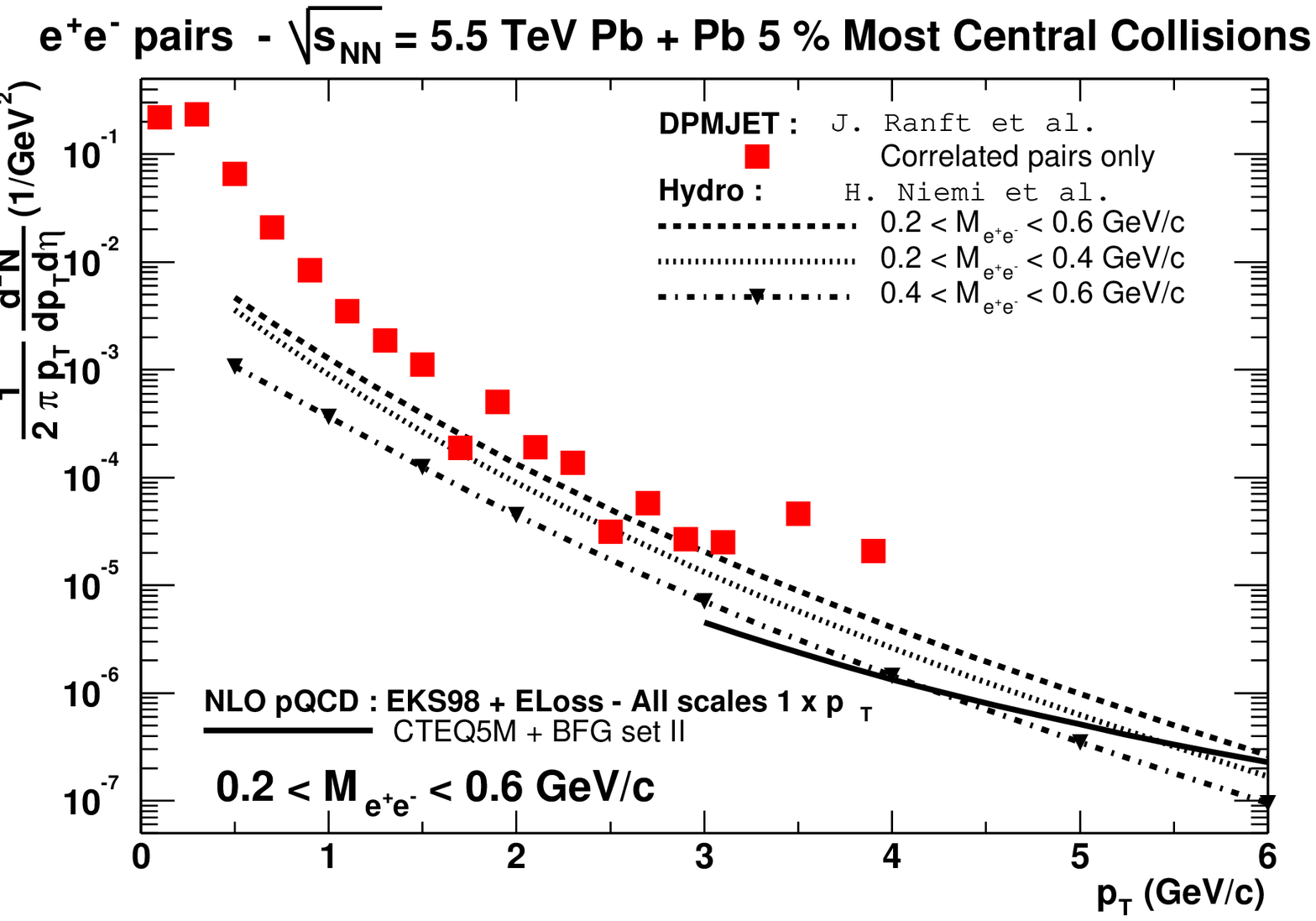}
\end{center}
\caption{The various components of the single dilepton $p_T$ spectrum in $Pb
Pb$ collisions at LHC, for small dilepton masses. The squares are an estimate of
the background.}
\label{figure5}
\end{figure}
The model being constrained by present data, it is justified to attempt
predictions for LHC (fig.~\ref{figure4}). Thermal photons dominate over
prompt photons up to rather large values of $p_T$, namely 6 GeV. It should
stressed that thermal photons are predominantly produced in the QGP phase while
at RHIC the hadronic phase is dominant. For real photons the background from
hadronic decays is very large in the low $p_T$ region. This is to be
constrasted with model estimates of the dilepton spectrum ($.2 < M_{e^-e^+}$
[GeV] $<.8$) where the background becomes of the same order as the thermal rate
already at 3 GeV (fig.~\ref{figure5}). Low mass dileptons appear therefore as an important
observable to directly probe the thermal effects. This has been illustrated in
the case of RHIC by the recent measurement of low mass dileptons by
PHENIX~\cite{phenix-dilep} in the low $p_T$ region inaccessible to real
photons because of the overwhelming background.

The last class of photon emission processes is the jet conversion mechanism
where a hard quark (anti-quark) radiates a photon in a Compton or annihilation
process with a thermal gluon or anti-quark (quark)~\cite{fries}. The resulting
spectrum is power behaved, reflecting the spectrum of the initial hard quark
and it should contribute in the intermediate $p_T$ region. These processes are
discussed by C.~Gale and S.~Jeon~\cite{photons,talks}. The calculation of their rate
requires modeling the space time evolution of the hard quark in the medium and
therefore the model dependence is expected to be rather large. The fact that
present data are consistent with models without jet conversion~\cite{yellowrep}
or with jet conversion~\cite{photons,talks} gives an indication on the theoretical
uncertainties.

In conclusion, the single real or virtual photon spectra in a heavy ion
collision receive contributions from several mechanisms: at the lowest $p_T$,
thermal effects dominate the signal; at medium $p_T$ jet conversion may play a
role while, at the upper end of the spectrum, the medium modified
bremsstrahlung and direct processes dominate. Turning to RHIC data, it is not
yet possible to constrain experimentally the relative normalisation of each
class of processes. Other observables should be considered to distinguish the
various mechanisms.

One such class of observables concerns correlation functions involving a
photon. One selects a large $p_T$ photon, to insure that it is produced in a
prompt process (the largest the transverse momentum, the smallest the 
bremsstrahlung component) and one can construct various measures involving this
photon and hadrons or photons found in the decay of the recoiling
jet~\cite{arleo}.  
\begin{figure} 
\begin{center}
\includegraphics[width=14cm,height=9cm]{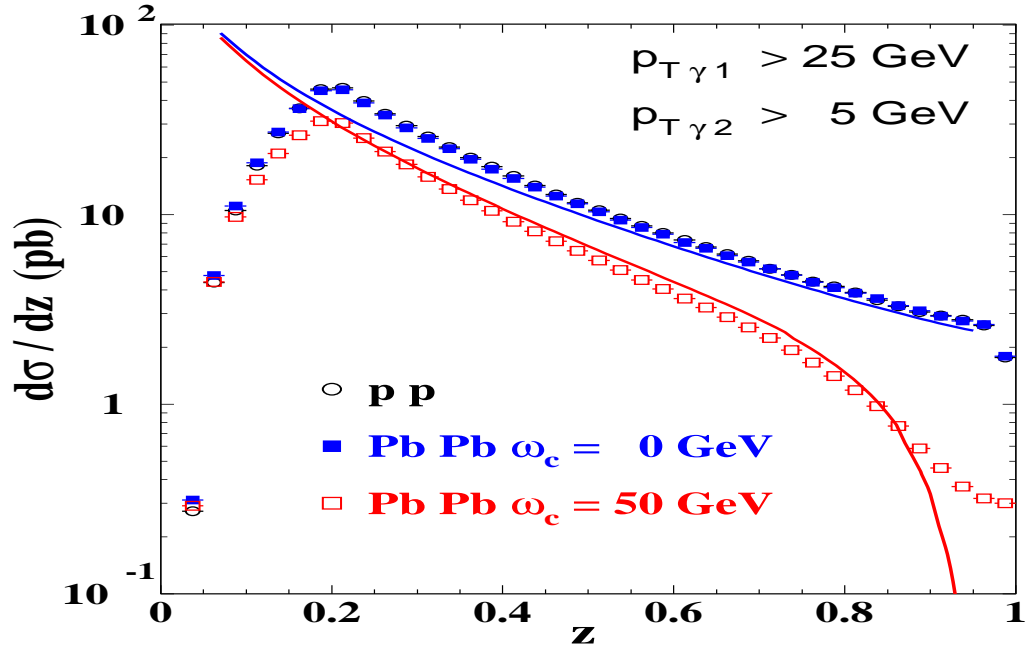}
\end{center} 
\caption{Exemple of a photon-photon correlation function at LHC ({\em courtesy 
F. Arleo}).} 
\label{figure6} 
\end{figure} 
Comparing these measures
in $pp$ and $AA$ collisions should give a direct information on the medium
modifications to the jet fragmentation function. Of particular interest is the
variable $z = -  {\bf p}^\gamma_{T} \cdot {\bf p}^a_{T} / {\bf p}^{\gamma
2}_{T} $ where ${\bf p}^a_{T}$ is the transverse momentum of a hadron or a
photon recoiling from the large $p_T$ photon. In a leading order calculation, if the hard
photon is directly produced, one has simply $z \sim z_{\rm frag}$.  An
illustration is shown in fig.~\ref{figure6} with the $z$ distribution for a
photon pair: the curve labeled $\omega_c = 0$ GeV makes use of the parton into
photon fragmentation function as in the vacuum, while the curve $\omega_c = 50$
GeV is for a model energy loss mechanism appropriate for LHC. Surperimposed on
the correlation curves are the input fragmentation functions. The $z$
distributions appear to follow closely the input functions
and therefore provide a way to probe in detail the jet energy loss mechanism.
The same measure involving a pion (photon-pion correlation) would have a larger
rate but leads to a more complex picture because of the convolution with the
production processes~\cite{arleo}. At small $z$ the dominant process is direct photon while at
large $z$ it is bremsstrahlung production where the relation $z \sim z_{\rm
frag}$ does not hold. The simple relation between the fragmentation function and
the observable is lost. However, comparing the $pp$ and $AA$ cases 
gives important information on the fragmentation process.

In conclusion, observables involving photons provide promising signals for
thermal effects. Due to the variety of emission mechanisms single spectra do
not allow to disentangle between the various channels. Correlation observables
such as that presented above provide detailed information on the hard parton
interaction in the hot medium.

I thank F.~Arleo for discussions and H.~Delagrange for discussions and providing
many figures.
I also thank the organizers of the meeting for financial support.

\end{document}